# A Robust Cross-Domain IDS using BiGRU-LSTM-Attention for Medical and Industrial IoT Security

Afrah Gueriani[a,c], Hamza Kheddar[a,c], Ahmed Cherif Mazari[a] and Mohamed Chahine Ghanem[*,b,c]

[a]*Laboratory of Advanced Electronic Systems (LSEA), University of Medea, Medea 26000, Algeria*
[b]*Cybersecurity Institute, Department of Computer Science, University of Liverpool, Liverpool, L693BX, UK*
[c]*Cyber Security Research Centre, London Metropolitan University, London, N78DB, UK*

## ARTICLE INFO

## ABSTRACT

The increased Internet of Medical Things (IoMT) and the Industrial Internet of Things (IIoT) interconnectivity has introduced complex cybersecurity challenges, exposing sensitive data, patient safety, and industrial operations to advanced cyber threats. To mitigate these risks, this paper introduces a novel transformer-based intrusion detection system (IDS), termed BiGAT-ID—a hybrid model that combines bidirectional gated recurrent units (BiGRU), long short-term memory (LSTM) networks, and multi-head attention (MHA). The proposed architecture is designed to effectively capture bidirectional temporal dependencies, model sequential patterns, and enhance contextual feature representation. Extensive experiments on two benchmark datasets; CICIoMT2024 (medical IoT) and EdgeIIoTset (industrial IoT); demonstrate the model's cross-domain robustness, achieving detection accuracies of 99.13% and 99.34%, respectively. Additionally, the model exhibits exceptional runtime efficiency, with inference times as low as 0.0002 seconds per instance in IoMT and 0.0001 seconds in IIoT scenarios. Coupled with a low false positive rate, BiGAT-ID proves to be a reliable and efficient IDS for deployment in real-world heterogeneous IoT environments.

## 1. Introduction

The rapid expansion of Internet of Things (IoT) technologies has significantly transformed various sectors, including healthcare, industry, agriculture, transportation, and smart cities, by offering enhanced services and improving operational efficiency [1]. With the continuous advancement of technology, the number of connected devices has grown substantially. This growth is largely driven by the integration of IoT and its associated communication systems, which now play a crucial role in both everyday applications and critical infrastructures across multiple domains. Within the healthcare sector, a specialized branch of IoT—commonly referred to as the Internet of Medical Things (IoMT), smart healthcare, or Healthcare 4.0, has emerged as a transformative force. IoMT enables continuous, real-time monitoring through connected medical devices, supporting applications such as telemedicine, wearable sensors, and remote diagnostics. These technologies have significantly improved patient care by enabling proactive intervention and seamless healthcare delivery.

In both IIoT and IoMT, the system architecture is typically structured into three primary layers: device layer, network layer, and application layer [2]. The *device layer* (or perception layer) comprises various sensors and smart devices responsible for data collection and environmental sensing. In IoMT, this includes patient monitoring devices, implantable sensors, and telemedicine equipment, while in IIoT it involves sensors from smart homes, vehicles, and industrial equipment [2]. The *network layer* ensures secure and efficient data transmission between devices and platforms using protocols specific to each domain. In both IIoT and IoMT environments, communication relies on lightweight and domain-specific protocols designed to support interoperability, real-time monitoring, and system control. Typical protocols include Modbus, MQTT, and LoRaWAN in industrial applications, as well as Bluetooth low energy (BLE), Zigbee, and Wi-Fi in healthcare systems. These communication technologies play a critical role in enabling seamless data exchange between devices and platforms, facilitating remote monitoring, and supporting the deployment of digital twins for smart factories, smart hospitals, and intelligent building management systems. At the top of the architecture, the *application layer* in both IIoT and IoMT systems enables intelligent decision-making, automation, and user interaction [3, 4]. In healthcare, this layer supports functionalities such as electronic health records, clinical decision support systems, and remote patient monitoring platforms [5]. Similarly, in industrial environments, the application layer integrates supervisory control and data acquisition (SCADA) systems, human-machine interfaces (HMI), and tools for real-time industrial process monitoring [6]. Across both domains, this layer transforms raw sensor data into actionable insights, supporting the deployment of digital twins and enabling proactive, data-driven responses to dynamic environments [7, 8].

### 1.1. Motivation and research gap

IoMT and IIoT technologies have enhanced interconnectivity and operational efficiency; however, this growing reliance on interconnected networks has also heightened their vulnerability to sophisticated and targeted cyber-attacks. Each layer of the IIoT and IoMT architecture presents specific security challenges and is exposed to various cyber threats. The *device layer* is vulnerable to physical attacks such as node tampering, fake node injection, replay attacks, and eavesdropping, targeting sensor data integrity and device functionality. The *network layer* is a primary target for communication-based attacks, including denial-of-service (DoS), distributed DoS (DDoS), man-in-the-middle (MITM) attacks, and unauthorized access, aiming to disrupt data transmission and network availability [9]. Meanwhile, the *application layer* faces software-level threats such as malware, ransomware, SQL injection, and privacy breaches, which can compromise sensitive data, critical services, and decision-making processes[10, 11].

*Dr Mohamed Chahine Ghanem is the corresponding author.
✉ gueriani.afrah@univ-medea.dz (A. Gueriani);
kheddar.hamza@univ-medea.dz (H. Kheddar);
mazari.ahmedcherif@univ-medea.dz (A.C. Mazari);
mohamed.chahine.ghanem@liverpool.ac.uk (M.C. Ghanem*)
ORCID(s):

Motivated by these critical needs, intrusion detection system (IDS) solutions have been developed for industrial control environments, including control network analysis, protocol analysis, and traffic mining based on machine learning or deep learning (DL) techniques [12]. However, traditional IDS approaches often fall short in addressing the increasing complexity and sophistication of modern cyber-attacks. This limitation is particularly critical in sensitive environments such as healthcare and industry, where security breaches may result in severe consequences, including data leakage, service disruption, operational downtime, and safety risks. These challenges underscore the urgent need for advanced IDS frameworks capable of real-time threat detection and rapid response, in order to safeguard both IoMT and IIoT networks against emerging and evolving threats [13].

### 1.2. Our contribution

This study proposes a hybrid DL-based IDS, named bidirectional GRU and Attention-based Transformer for intrusion detection *(BiGAT-ID)*. Within the proposed IDS, IoT traffic is modeled as a *non-stationary* sequence where attack semantics often depend on short transients (e.g., flag flips, rapid port scans) embedded in longer session patterns. To address this, a BiGRU encoder is employed for efficient *past–future context capture*, while an LSTM layer ensures *long-range* attack persistence through stable gradient flow that mitigates vanishing gradients. To enhance discrimination, a multi-head attention (MHA) mechanism is applied, improving detection accuracy and reducing false positives by *focus on salient patterns* within the traffic sequence [14]. The main contributions of this study are summarized as follows:

- To the best of the authors' knowledge, BiGAT-IDS uniquely combines BiGRU, LSTM, and MHA to achieve outstanding intrusion detection performance.
- BiGAT-IDS detects attacks across IIoT and IoMT domains, proving its versatility and strong generalization.
- Leave-one-attack-out (LOAO) zero-day strategy testing confirms that BiGAT-IDS detects both signature-based and zero-day threats across distinct attack patterns.
- BiGAT-IDS outperforms state-of-the-art models on both datasets, validating its superior detection accuracy.

## 2. Related works

Numerous techniques have been developed to combat intrusions in IoMT networks. For example, the work in [15] introduces a CNN-based IDS for IoMT networks that outperforms traditional ML methods in accuracy and efficiency, demonstrating strong robustness and real-world applicability for healthcare cybersecurity. Moreover, both [17] and [16] utilize deep neural networks (DNN) for intrusion detection but differ in focus and design. It integrates DNN within a federated learning framework for decentralized IoT, emphasizing scalability and privacy, while [17] employs a self-attention-based DL model for IoMT anomaly detection. Notably, [16] outperforms [17] in terms of F1-score, demonstrating superior effectiveness across the same dataset. LSTM-based approaches form the foundation of several studies [26, 27]. The DAG-LSTM model [28] enhances intrusion detection in IoMT by incorporating feature optimization techniques, achieving an accuracy of 92%. In contrast, the EFL-LSTM model [29] leverages federated learning to enable decentralized detection, reaching a higher accuracy of 97.16%. Both schemes are evaluated solely on the ECU-IoHT benchmark dataset, which limits their ability to detect a wide range of unseen IoT attacks. Similarly, an advanced LSTM-based intrusion detection scheme is proposed in [18] to secure IoMT environments. By learning from diverse equipment and attack patterns, the model effectively captures complex threat behaviors to safeguard sensitive medical data against evolving cybersecurity challenges. However, the scheme has not been evaluated against IIoT-specific attacks. Stacking ensemble models combining LSTM, CNN, and DNN in [30] achieve high performance on the ECU-IoHT dataset for IoMT detection. However, the approach incurs high computational cost and lacks evaluation on IIoT or general IoT datasets, raising concerns about its generalizability and real-world adaptability. Innovative architectures include CKANs, which embed the Kolmogorov-Arnold theorem with self-attention mechanisms to balance accuracy and interpretability in IoMT, though with trade-offs in resource efficiency [31]. Yet, CKANs aim to balance interpretability and accuracy; this trade-off might not hold across all datasets or intrusion types, leading to potential performance drops. Meanwhile, FlowID [19] leverages a hypergraph-based DL framework combined with dual-contrastive self-supervised learning to effectively capture high-order traffic interactions. This design enhances its robustness and generalization across diverse network scenarios and benchmark datasets.

Researchers have shown increasing interest in developing techniques to counter intrusions in IIoT networks compared to IoMT. For example, in [20] and [21], CNN-LSTM architectures are proposed for IIoT intrusion detection. The first employs LSTM only, while the second incorporates a self-attention layer to enhance accuracy. Both models are evaluated on the Edge-IIoTset dataset, demonstrating strong performance in binary and multiclass classification. The self-attention-based model achieves superior accuracy and low inference time, making it well-suited for real-time and robust IIoT cybersecurity applications. The simplest approach is observed in [32], which explores standard DL architectures such as artificial neural networks (ANN), LSTM, and GRU for attack classification. These models are applied independently and lack attention for contextual focus. A step further, [22] employs a dual recurrent approach by combining Bi-GRU with LSTM to improve the modeling of sequential dependencies in network traffic. This fusion enhances the learning of long-range temporal features while maintaining a focus on recurrent architectures. More advanced still, [33] proposes the Magru-IDS model, which integrates GRU with MHA to refine the extraction of relevant temporal features. This addition of attention mechanisms improves the model's ability to selectively focus on critical

**Table 1**
Comparative summary of state-of-the-art IDS solutions for IoMT and IIoT networks.

|  | Ref | Dataset | DL method | BP (%) | Limitations | Bi/GRU | Employing LSTM | MHA |
|---|---|---|---|---|---|---|---|---|
| Medical domain | [15] | CICIoMT2024 | CNN | Acc=99.00 F1= 98.00 | Absence of comparison with prior works, and evaluation conducted on a single dataset | ✗ | ✗ | ✗ |
| | [16] | CICIoMT2024 | DNN | Acc=99.56 Pr=94.59 | Lacks temporal awareness and generalizes poorly on sequences, and use imbalanced dataset | ✗ | ✗ | ✗ |
| | [17] | CICIoMT2024 | Att-DNN | Pr=84.43 F1= 91.02 | Protocol diversity and data imbalance hinder robust cross-domain model generalization. | ✗ | ✗ | ✓ |
| | [18] | CICIoMT2024 | L2D2 | Acc=98.00, F1=98.00 | Not evaluated for delays, zero-day threats, or IIoT generalization. | ✗ | ✓ | ✗ |
| | [19] | CIC-IOMT2024 | CNN-GNN | Acc=95.27 F1 = 90.36 | High complexity limits deployment on resource-constrained devices. | ✓ | ✗ | ✗ |
| Industrial domain | [16] | CICIoT2023 | DNN | Acc=99.09 Pr=91.56 | Lacks temporal awareness and generalizes poorly on sequences, and use imbalanced dataset | ✗ | ✗ | ✗ |
| | [17] | MQTT-IoT-IDS | Att-DNN | Pr=92.14 F1=95.53 | Protocol diversity and data imbalance hinder robust cross-domain model generalization. | ✗ | ✗ | ✓ |
| | [20] | EdgeIIoTset | CNN-LSTM | Acc=98.68 | Evaluation based on limited metrics and a single dataset affects overall generalizability. | ✗ | ✓ | ✗ |
| | [21] | EdgeIIoTset | LSTM-CNN-Att | F1=99.04, FPR=0.002 | Lacks mechanism to emphasize key temporal patterns and cross-domain IIoT–IoMT applicability. | ✗ | ✓ | ✓ |
| | [22] | Edge-IIoTset | BiGRU-LSTM | Acc=98.32, FPR=0.046 | Lacks attention mechanism to highlight important temporal features and patterns. | ✓ | ✓ | ✗ |
| | [23] | EdgeIIoTset | CNN-GRU | Acc= 98.70 FPR=0.7 | Valid for single dataset, no advanced comparison, and high FPR. | ✓ | ✗ | ✗ |
| | [24] | EdgeIIoTset | CNN-LSTM-GRU | Acc= 97.44 | Limited metrics, single unbalanced dataset, no dynamic focus on temporal patterns. | ✓ | ✓ | ✗ |
| | [25] | Edge-IIoTset | BiGRU-CNN | Acc=94.7 Pr=94.8 | Limits capturing long-range dependencies, weakening performance on extended or noisy sequences. | ✓ | ✗ | ✓ |

Abbreviations: Best performance (BP); Attention (Att)

time-dependent patterns. However, the model has limited sequential dependency modeling capability. Another strategy, like [34] introduces a scheme that augments a conventional CNN by capturing global feature dependencies through self-attention. This work also applies data cleaning and mutual information-based feature filtering, enhancing the model's feature selection without adding sequence modeling. However, the scheme suffer from weak temporal sequence learning ability. Increasing in complexity. Similarly, in [23], which combines CNN and GRU. The CNN component captures spatial patterns, while GRU processes temporal sequences, allowing the model to handle both static and dynamic data characteristics effectively. A more complex hybrid strategy is presented in [24], where CNN, LSTM, and GRU are combined into a triple-ensemble model. This structure leverages the individual strengths of each DL block, spatial, sequential, and gated temporal processing, resulting in a more expressive and robust intrusion detection framework. Finally, the most sophisticated architecture appears in [25], which integrates self-attention mechanisms, BiGRU, and Inception-CNN. Beyond model complexity, this work incorporates advanced data handling techniques, including mixed sampling, denoising, and feature selection based on Pearson correlation and Random Forest. According to Table 1, many existing IDS models demonstrate strong performance but suffer key limitations. Most fail to integrate BiGRU, LSTM, and MHA in a unified framework, reducing their capacity to capture comprehensive temporal and contextual features. Additionally, many are domain-specific, limiting cross-environment generalization between IoMT and IIoT systems. Crucial deployment aspects such as inference time and runtime efficiency are often neglected, undermining real-world applicability. Finally, explainability is rarely addressed, despite its importance for transparency and trust in critical settings like healthcare. These limitations motivate the development of more holistic, generalizable, and interpretable IDS frameworks.

## 3. Proposed Methodology

The proposed model synergistically combines BiGRU for capturing forward-backward dependencies, LSTM for deep sequential feature extraction, and MHA for extracting long dependencies salient temporal features. Unlike existing models evaluated on single-domain datasets with shallow temporal encoding, BiGAT-ID enables robust intrusion detection across both IoMT and IIoT environments. This composite architecture enhances model interpretability, minimizes false alarms, and maintains low inference latency, ensuring scalability and real-time applicability in heterogeneous operational settings. The following subsections detail the preprocessing pipeline and architectural components. The architecture of the proposed BiGAT-ID, illustrated in Figure 1, is organized into two main components: the data preprocessing pipeline and the model architecture.

### 3.1. Pre-processing

The preprocessing stage plays a crucial role in converting raw, heterogeneous network traffic data into a structured format suitable for DL-based IDS. The process typically involves several essential steps:

- **Data cleaning and label encoding:** Dataset structure and labels are analyzed; categorical fields are numerically encoded using LabelEncoder for model compatibility.
- **Feature extraction and reshaping:** Selected features are reshaped into 3D matrices to align with sequential model input and capture temporal dependencies.

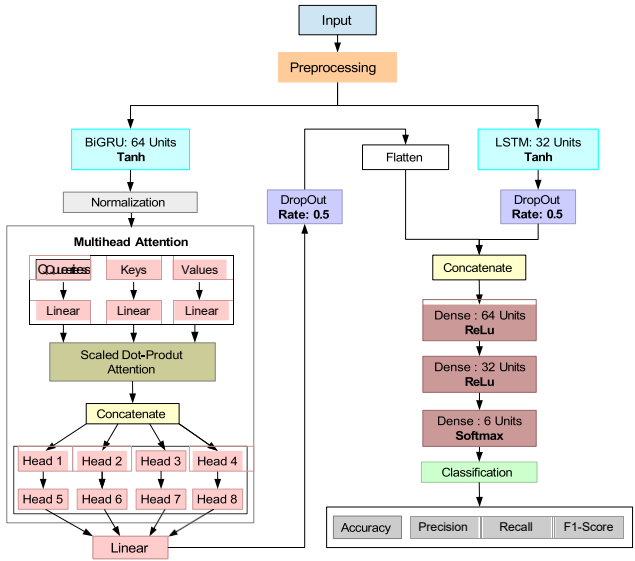

**Figure 1:** The proposed BiGAT-IDS model.

**Table 2**
BiGAT-ID model summary and hyperparameters.

|  | DL Layer | Unit | Output Shape | Connected to |
|---|---|---|---|---|
| 1 | Input | – | IoMT: (None, 83, 1) IIoT: (None, 60, 1) | – |
| 2 | BiGRU | 64 | (None, 83, 128) | Input_Layer[0][0] |
| 3 | LayerNorm. | - | (None, 83, 128) | BiGRU |
| 4 | MHA | - | (None, 83, 128) | LayerNorm. |
| 5 | Dropout_1 | - | (None, 83, 128) | MHA (8, 64) |
| 6 | Flatten | - | (None, 10624) | Dropout_1 |
| 7 | LSTM | 32 | (None, 32) | Input_Layer[0][0] |
| 8 | Dropout_2 | - | (None, 32) | LSTM |
| 9 | Concatenate | - | (None, 10656) | Flatten, Dropout_2 |
| 10 | Dense | - | (None, 64) | Concatenate |
| 11 | Dense | - | (None, 32) | Dense |
| 12 | Dense | - | (None, 6) | Dense |
| Total parameters: | | | 978,470 | |
| Trainable parameters: | | | 978,470 | |
| Non-trainable parameters: | | | 0 | |

- **Class imbalance mitigation:** Random over sampler (RoS) addresses IoMT sparsity by duplicating existing minority samples, while SMOTE enhances IIoT class balance by synthetically generating new minority samples through feature-space interpolation, improving generalization compared to simple duplication [35]. Focal loss was further incorporated during training to down-weight well-classified examples and emphasize hard and minority-class samples, thereby improving sensitivity to subtle, rare attack patterns [36].
- **Dataset splitting:** Data is split into 80% training and 20% testing sets using stratified sampling via `train_test_split` for balance.
- **Label transformation:** Response labels are one-hot encoded using `to_categorical` to support categorical cross-entropy loss in multiclass classification.

### 3.2. Model architecture

The BiGAT-ID model processes sequential data shaped as (83,1) for IoMT and (60,1) for IIoT. Its dual-branch architecture (Figure 1) captures complementary features. Table 2 outlines the model structure and hyperparameters.

*- Branch 1: BiGRU with MHA:* This branch uses a BiGRU layer (64 units) followed by Layer Normalization, multi-head attention (8 heads, 64 key dim), and Dropout (0.5). It captures bidirectional dependencies and highlights salient temporal patterns. Output is flattened for fusion.

*- Branch 2: LSTM with Dropout:* A single LSTM layer (32 units, `return_sequences=False`) extracts condensed temporal features, followed by Dropout (0.5). This setup ensures a compact embedding of input sequences suitable for classification while reducing overfitting through neuron deactivation during training.

*- Fusion and classification:* Outputs from both branches are concatenated and passed through Dense layers (64, 32 units with ReLU). A final Dense layer (6 units, softmax) generates class probabilities for multiclass classification, leveraging learned features from both processing paths.

*- Model compilation and validation:* The model is compiled with `Adam` optimizer and categorical cross-entropy loss. Performance is evaluated using accuracy, FPR, precision, recall, F1-score, and inference time [37], ensuring robust validation of classification effectiveness.

## 4. Experimentation

### 4.1. Datasets

This study employs two publicly available benchmark datasets to evaluate the proposed BiGAT-ID model, summarized in Table 3.

1. **CICIoMT2024**[1]: Designed to strengthen IoMT security research, this dataset includes 18 attack types grouped into five (5) categories, captured from a testbed of 40 devices (25 physical and 15 simulated). It features key IoMT protocols such as Wi-Fi, MQTT, and Bluetooth [38].

2. **Edge-IIoTset**[2]: Collected from a 7-layer IIoT testbed with 10 smart devices, this dataset contains 14 attack types across six (6) categories. It focuses on IIoT-specific traffic and protocols, using 61 selected features out of 1,176 [39].

3. **TON_IoT**[3]: Encompasses heterogeneous data sources derived from telemetry records of IoT/IIoT services, operating system logs, and network traffic within an IoT environment. It comprises 44 features and incorporates benign, and nine (9) distinct categories of cyberattacks, namely scanning, DoS, DDoS, ransomware, backdoor, data injection, XSS, password cracking, and MITM; executed against various IoT and IIoT sensors across the IIoT network, captured from a testbed of more than 10 devices. This dataset is publicly accessible through the TON_IoT repository [40].

---
[1] https://www.unb.ca/cic/datasets/tabular-iot-attack-2024.html
[2] https://www.kaggle.com/datasets/cnrieiit/mqttset
[3] https://www.kaggle.com/datasets/alaaelmor/ton-iot-train-test-network

Table 3
Datasets utilized in the study, along with their respective attack categories, types, and the number of samples.

| Data | Category | Samples | Attack types |
|---|---|---|---|
| CICIoMT2024 | Normal | 32620 | – |
| | DDoS | 2576 | DDoS SYN, DDoS TCP, DDoS ICMP, DDoS UDP |
| | DoS UDP | 3115 | DoS SYN, DoS TCP, DoS ICMP, DoS UDP |
| | MITM | 1053 | ARP Spoofing |
| | MQTT | 953 | Malformed Data, DoS Connected Flood, DDoS Publish Flood, DoS Publish Flood |
| | Recon | 8321 | Ping Sweep, Recon VulScan, OS Scan, Port Scan |
| EdgeIIoTset | Normal | 24301 | – |
| | DDoS | 49396 | DDoS_UDP, DDoS_TCP, DDoS_ICMP, DDoS_HTTP, DDoS_HTTP |
| | Info gathering | 21148 | Port Scanning, Fingerprinting, Vulnerability_Scanner |
| | MITM | 1214 | DNS Spoofing, ARP Spoofing |
| | Injection | 30632 | SQL_Injection, XSS, UpLoading |
| | Malware | 31109 | Ransomeware, Backdoor, Password |

### 4.2. Obtained results

- **Accuracy and loss graph:** Figure 2 illustrates the training and validation curves for the multiclass attack detection model. Subfigure (a) depicts the accuracy trends, while subfigure (b) shows the loss trajectories across epochs for both datasets.

- **Accuracy curves (Figure 2(a)):** For the CICIoMT2024 dataset, the training accuracy rapidly approaches 100%, and the validation accuracy follows closely, stabilizing at 99.13%, which indicates excellent generalization capability. For the EdgeIIoTset dataset, both training and validation accuracy curves follow a consistently upward trajectory with minor oscillations, achieving a peak validation accuracy of 99.34%. These results confirm the model's strong classification ability in both medical and industrial IoT environments.

- **Loss curves (Figure 2(b)):** The training and validation loss curves for CICIoMT2024 demonstrate a steady decline, converging to a minimal loss value of 0.0257%. A similar downward trend is observed for the EdgeIIoTset dataset, where the final validation loss reaches 0.0158%. The consistent reduction in loss across both domains reflects effective learning and minimal overfitting.

- *Classification report analysis:* Table 4 presents a detailed classification report for the proposed BiGAT-ID model on both CICIoMT2024 and EdgeIIoTset datasets, reflecting its robust performance across multiple cyberattack classes.

- **IoMT domain:** BiGAT-ID achieves 100% F1-scores in 4 of 6 classes, and 98% in others. With 39,144 validated samples, the model shows excellent precision and recall, effectively distinguishing normal and malicious IoMT traffic.

- **IIoT domain:** On 59,276 validated samples, the model scores 100% F1 in MITM and Malware, and 97–99% in other attacks. BiGAT-ID maintains 99% accuracy, confirming its robust detection capability across diverse IIoT attack types.

Table 4
Classification report for the proposed BiGAT-ID model on CICIoMT2024 and EdgeIIoTset datasets in percentage (%).

| CICIoMT2024 dataset | | | | EdgeIIoTset dataset | | | |
|---|---|---|---|---|---|---|---|
| Attack type | Pr | Rc | F1 | Attack type | Pr | Rc | F1 |
| Normal | 98 | 98 | 98 | Normal | 100 | 100 | 100 |
| DDoS UDP | 100 | 100 | 100 | DDoS | 98 | 96 | 97 |
| DoS UDP | 97 | 98 | 98 | Info. gath. | 98 | 99 | 99 |
| MITM | 100 | 100 | 100 | MITM | 100 | 100 | 100 |
| MQTT | 100 | 100 | 100 | Injection | 97 | 98 | 98 |
| Recon. | 100 | 100 | 100 | Malware | 100 | 100 | 100 |
| Accuracy | | | 99 | Accuracy | | | 99 |
| Macro avg | 99 | 99 | 99 | Macro avg | 98 | 98 | 98 |
| Weighted avg | 99 | 99 | 99 | Weighted avg | 99 | 99 | 99 |

Abbreviation: Validation data (VD).

Macro and weighted F1-scores between 98–99% confirm BiGAT-ID's robustness against class imbalance and diverse attack types.

- *Confusion Matrix:* Figure 3 presents the normalized confusion matrices for the CICIoMT2024 and EdgeIIoTset datasets, shown in subfigures (a) and (b), respectively. These matrices provide detailed insights into the classification performance of the BiGAT-ID model across all classes.

- In Figure 3 (a), which presents results for the CICIoMT2024 dataset, the model achieves high accuracy across all attack types. Benign traffic attains a TPR of 98% with only 2% misclassified as MQTT, indicating strong separation between benign and malicious flows. MITM achieves a perfect TPR of 100%, reflecting excellent pattern capture. MQTT records a TPR of 97% but 2% are misclassified as benign, raising concern over false negatives and suggesting the need for improved feature extraction for this attack type. Recon, DDoS, and DoS all achieve a perfect TPR of 100%, showing error-free detection.

- In Figure 3 (b), which presents results for the EdgeIIoTset dataset, the model maintains strong classification performance. Normal traffic achieves a TPR of 100% with no misclassifications, indicating highly reliable benign traffic detection. The DDoS attack is classified with 99% accuracy, with 1% misclassified into the injection class. The injection class shows a slight drop in precision, with 3% of instances misclassified as DDoS, resulting in 97% accuracy. This minor weakness may stem from shared characteristics such as abnormal request rates between injection and volumetric attacks like DDoS. MITM, information gathering, and malware are classified with perfect accuracy of 100%, underscoring the model's precision in detecting critical malicious traffic patterns and confirming its robustness.

The minimal confusion observed between certain traffic types reflects realistic challenges in intrusion detection while still showcasing high generalization and discriminative power across diverse IoT/IIoT attack surfaces, with confusion in CICIoMT2024 limited to benign–MQTT cases and in EdgeIIoTset limited to the injection class. These confusions are minimal and acceptable given the complexity of multiclass classification in a medical IoT environment. The BiGAT-ID model demonstrates highly accurate detection on CICIoMT2024, especially for MITM, Recon, DDoS,

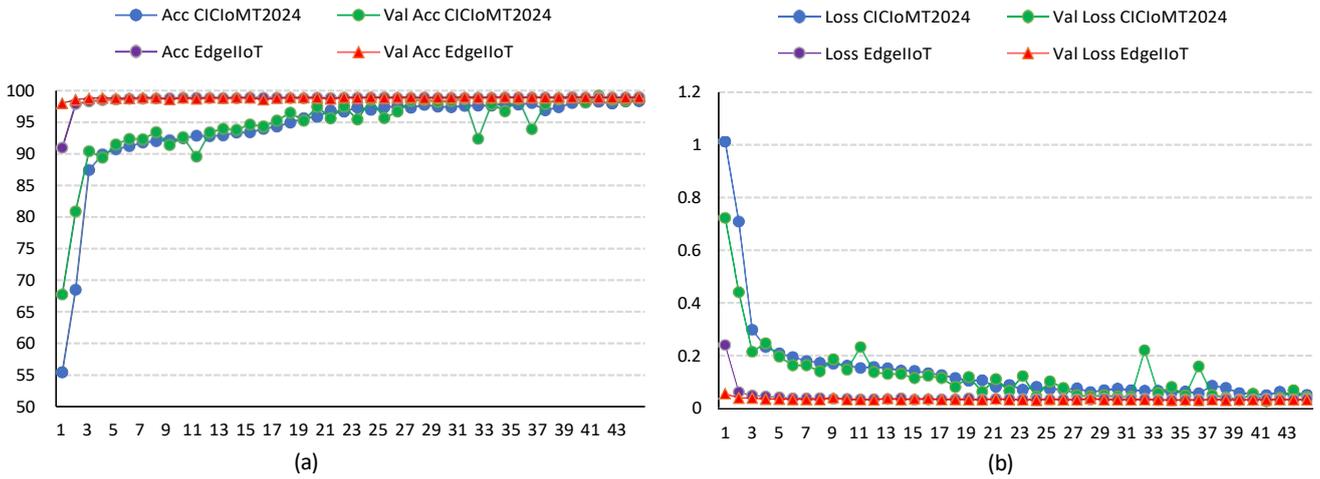

**Figure 2:** Accuracy and loss metrics for the proposed BiGAT-ID model across both datasets. (a): Accuracy, (b): Loss.

and DoS, and performs excellently across the EdgeIIoTset dataset.

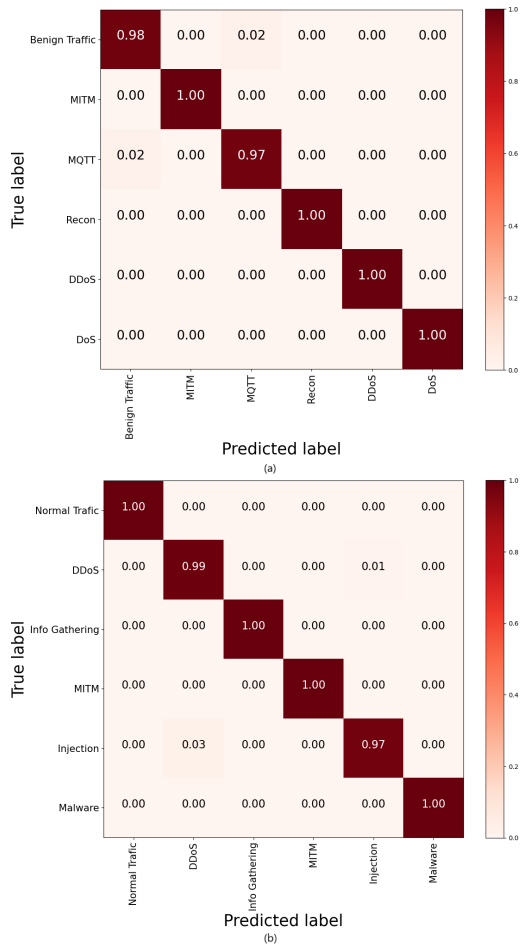

**Figure 3:** The confusion matrix of the BiGAT-ID model: (a) For the CICIoMT2024 dataset; (b) For the EdgeIIoTset dataset.

Figure 4 illustrates the ROC curves, derived from the confusion matrix, for multi-class classification performance on two datasets. In both subfigures, the curves show almost perfect separation for all six attack classes, tightly concentrated in the top-left corner, with each class achieving an AUC of 1.00. In Figure 4 (a) for the CICIoMT2024 dataset, this performance reflects ideal sensitivity and specificity across all classes, underscoring the model's robustness in complex medical IoT scenarios where precise detection of various attack types or system states is critical. The sharp curve rise toward the top-left further indicates an extremely low false positive rate (FPRs) even at high TPRs. In Figure 4 (b), for the Edge-IIoT dataset, the uniform ROC curves across classes confirm strong generalization and the absence of class bias, which is crucial in real-world intrusion detection. These results highlight both the architecture's discriminative strength and its ability to accurately identify diverse intrusion types.

To address potential overfitting risks associated with oversampling minority classes, multiple mitigation strategies were implemented. The datasets chosen were accurately labeled and consistently structured in terms of features, while rigorous preprocessing, tailored to each dataset, was applied, including data cleaning and class imbalance handling as described in Section 3. Furthermore, a learning rate of $10^{-3}$ and a dropout rate of $0.5$ were adopted to enhance generalization. Collectively, these measures ensured stable and robust performance, with no observable signs of overfitting throughout the experiments.

- *Inference time:* Inference time is vital for real-time IDS deployment. BiGAT-ID achieves impressive efficiency with 0.0002s per instance in IoMT and 0.0001s in IIoT environments. These results ensure rapid threat detection without latency, which is critical in healthcare systems to protect patient safety and in industrial systems to prevent costly disruptions. The low inference times highlight the model's suitability for real-world, resource-constrained settings where timely intrusion response is essential for operational continuity and data security.

### 4.3. Zero-day attack assessment

To further evaluate the generalization capability of the proposed model in detecting previously unseen cyber threats, we conducted a LOAO zero-day simulation across the datasets considered in our study. This experimental setup involved completely excluding a specific attack class during the training phase, while retaining all other classes. The proposed model demonstrated strong zero-day detection capabilities across all evaluated datasets. In our experiments, the model achieved a maximum validation accuracy of 98.50% with a corresponding validation loss of 0.04 on the EdgeIIoT dataset. For the CICIoMT2024 dataset, the proposed BiGAT-ID model attained a maximum validation accuracy of 99.23% with a loss of 0.01. On the TON_IoT dataset, the BiGAT-ID model reached a maximum validation accuracy of 98.07% with a loss of 0.06. These findings highlight the model's resilience in detecting unseen attack behaviors and underscore its effectiveness for real-world deployment scenarios where novel attacks are likely to emerge.

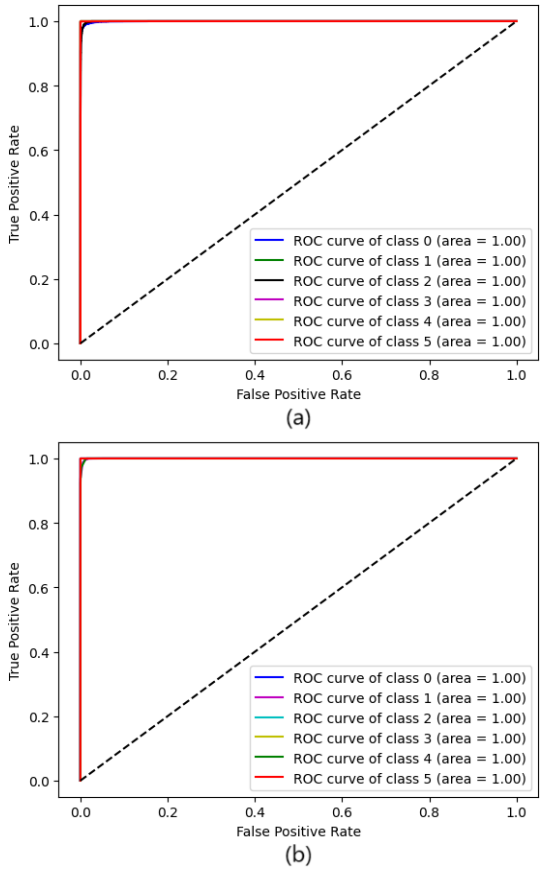

**Figure 4:** The ROC curves of the proposed BiGAT-ID model. (a) For the CICIoMT2024 dataset, (b) For the EdgeIIoTset dataset.

### 4.4. Ablation study

Table 5 presents the comparative performance of ten architectural variants of the proposed model across both CICIoMT2024 and EdgeIIoTset datasets, evaluated before and after data balancing. The core objective of this ablation study is to identify the configuration that maintains high performance when the model is deployed across both the medical (IoMT) and industrial (IIoT) domains, as well as consistency under data imbalance, a frequent characteristic in real-world intrusion detection scenarios.

The results reveal that configuration #4 (`BiGRU+MHA`)-LSTM, denoted as BiGAT-ID, consistently outperforms other variants. Specifically, before data balancing, it achieves an accuracy of **96.19%** (CICIoMT2024) and **99.04%** (EdgeIIoTset) with corresponding FPRs of **0.0211%** and **0.0013%**, respectively. After balancing, the same model further improves, reaching **99.13%** accuracy on CICIoMT2024 and **99.34%** on EdgeIIoTset, while maintaining an exceptionally low FPR of **0.0013%** for both datasets. Although other configurations, such as #3 (BiGRU-LSTM-MHA) and #6 (BiGRU-(MHA+LSTM)), also show strong results, they exhibit either higher FPRs or less stability under balancing conditions. For instance, configuration #3 suffers a noticeable drop in performance on CICIoMT2024 after balancing (from **95.30%** to **94.73%**) and shows a higher FPR of **0.0093%**, making it less optimal for deployment in sensitive domains like healthcare.

Furthermore, additional experiments (cases #7–#12) explored the impact of varying attention heads, dropout rates, and unit configurations. We empirically evaluated configurations with 2, 4, and 8 heads, finding that 8 heads offered the best balance between accuracy and computational efficiency. Increasing beyond this point yielded no significant gains but increased complexity. Dropout rates of 0.2, 0.3, 0.5, and 0.7 were tested, with 0.5 providing optimal regularization, effectively mitigating overfitting without inducing underfitting. Similarly, experiments with 64, 128, and 256 units showed that setting the LSTM to 32 units and the BiGRU to 64 units resulted in a balanced architecture that sustained strong performance while improving training efficiency.

### 4.5. BiGAT-ID explainability

The Shapley additive explanations (SHAP) summary plot in Figure 5 illustrates the average contribution of each feature to the model's output across IIoT and IoMT datasets. Feature importance is reflected by bar length, while colors denote class-specific impacts (class 0–5). The X-axis represents the mean SHAP magnitude, and the Y-axis lists the model features [41].

In Figure 5 (a), feature 6 *(Timestamp)* emerges as the most influential, with a SHAP value magnitude around 10, strongly impacting class 0 and class 2. feature 72 *(Bwd Init Win Bytes)* is particularly relevant for class 4, while feature 20 *(Flow Bytes/s)* and feature 4 *(Dst Port)* show moderate contributions across several classes. Other features, such as feature 30 *(Fwd IAT Min)*, feature 53 *(ACK Flag Count)*, and feature 43 *(Bwd Packets/s)*, provide smaller yet meaningful support, serving as secondary discriminators. The remaining features contribute minor but cumulative effects, likely refining predictions and reducing false positives through added contextual information.

In Figure 5 (b), feature 13 *(http.request.uri.query)* emerges as the most influential, particularly for class 4. Several features, such as feature 30 *(tcp.len)*, feature 10 *(icmp.unused)*, feature 14 *(http.request.method)*, and feature 31 *(tcp.options)*, also contribute significantly across multiple classes, while features 38 *(dns.qry.name)* and 44 *(dns.retransmit_request_in)* are dominant for benign traffic (class 0). Moderate features (e.g., *frame.time*, *ip.src_host*, *tcp.dstport*) strengthen predictions, whereas lower-impact features (e.g., *tcp.flags.ack*, *udp.port*) play supportive roles. Globally, the SHAP analysis highlights the interpretability of BiGAT-ID, confirming its ability to capture both global and class-specific patterns across CICIoMT2024 and EdgeIIoT datasets, without bias toward any single class.

### 4.6. Comparison with state-of-the-art

As summarised in Table 6, several recent IDS models demonstrate only moderate performance and are evaluated on a single dataset, limiting their generalizability.

**Table 5**
Performance of different variants of the proposed models in multiclass classification, with the endices indicating the hyperparameter values of each block.

| | | Before balancing data | | | | | |
|---|---|---|---|---|---|---|---|
| | | CICIoMT2024 | | | EdgeIIoTset | | |
| | Model | Acc. (%) | Loss (%) | FPR (%) | Acc. (%) | Loss (%) | FPR (%) |
| #1 | BiGRU$_{64}$+MHA$_8$ | 94.60 | 0.1804 | 0.0406 | 98.87 | 0.0364 | 0.0016 |
| #2 | LSTM$_{32}$+MHA$_8$ | 94.61 | 0.1884 | 0.038 | 98.33 | 0.0600 | 0.0023 |
| #3 | BiGRU$_{64}$-(LSTM$_{32}$+MHA$_8$) | 95.30 | 0.1665 | 0.0243 | 99.01 | 0.0284 | 0.0016 |
| **#4** | **(BiGRU$_{64}$+MHA$_8$)-LSTM$_{32}$** | 96.19 | 0.1368 | 0.0211 | 99.04 | 0.0262 | 0.0013 |
| #5 | (MHA$_8$+BiGRU$_{64}$)-LSTM$_{32}$ | 86.12 | 0.4893 | 0.0983 | 97.57 | 0.0841 | 0.0024 |
| #6 | BiGRU$_{64}$-(MHA$_8$+LSTM$_{32}$) | 95.88 | 0.1400 | 0.0200 | 99.11 | 0.0227 | 0.0013 |
| #7 | (BiGRU$_{128}$+MHA$_8$)-LSTM$_{256}$ | 94.62 | 0.1811 | 0.0456 | 97.73 | 0.0734 | 0.0079 |
| #8 | (BiGRU$_{64}$+MHA$_2$)-LSTM$_{32}$ | 94.21 | 0.1938 | 0.0478 | 98.55 | 0.0365 | 0.0038 |
| #9 | (BiGRU$_{64}$+MHA$_4$)-LSTM$_{32}$ | 94.41 | 0.1823 | 0.0356 | 98.85 | 0.0300 | 0.0178 |
| #10 | (BiGRU$_{64}$+MHA$_8$)-LSTM$_{32}$ "0.3 D" | 94.92 | 0.1688 | 0.0294 | 98.68 | 0.0522 | 0.0020 |
| #11 | (BiGRU$_{64}$+MHA$_8$)-LSTM$_{32}$ "0.7 D" | 93.96 | 0.1892 | 0.0486 | 99.00 | 0.0278 | 0.0016 |
| #12 | (BiGRU$_{64}$+MHA$_8$)-LSTM$_{32}$ "0.2 D" | 93.72 | 0.2068 | 0.0449 | 98.94 | 0.0303 | 0.0043 |
| | | After balancing data | | | | | |
| #1 | BiGRU$_{64}$+MHA$_8$ | 96.15 | 0.0828 | 0.007 | 99.32 | 0.0165 | 0.0013 |
| #2 | LSTM$_{32}$+MHA$_8$ | 98.41 | 0.0440 | 0.0026 | 98.88 | 0.0304 | 0.0023 |
| #3 | BiGRU$_{64}$-(LSTM$_{32}$+MHA$_8$) | 94.73 | 0.1100 | 0.0093 | 99.12 | 0.0194 | 0.0020 |
| **#4** | **(BiGRU$_{64}$+MHA$_8$)-LSTM$_{32}$** | 99.13 | 0.0257 | 0.0013 | 99.34 | 0.0158 | 0.0013 |
| #5 | (MHA$_8$+BiGRU$_{64}$)-LSTM$_{32}$ | 65.69 | 0.7986 | 0.0625 | 99.04 | 0.0260 | 0.0013 |
| #6 | BiGRU$_{64}$-(MHA$_8$+LSTM$_{32}$) | 93.74 | 0.1295 | 0.0106 | 99.32 | 0.0159 | 0.0013 |
| #7 | (BiGRU$_{128}$+MHA$_8$)-LSTM$_{256}$ | 90.74 | 0.2156 | 0.0323 | 60.27 | 1.1627 | 0.4231 |
| #8 | (BiGRU$_{64}$+MHA$_2$)-LSTM$_{32}$ | 93.54 | 0.1344 | 0.0146 | 99.21 | 0.0214 | 0.0016 |
| #9 | (BiGRU$_{64}$+MHA$_4$)-LSTM$_{32}$ | 95.64 | 0.0944 | 0.0090 | 99.26 | 0.0217 | 0.0013 |
| #10 | (BiGRU$_{64}$+MHA$_8$)-LSTM$_{32}$ "0.3 D" | 95.49 | 0.0978 | 0.0089 | 99.30 | 0.0230 | 0.0013 |
| #11 | (BiGRU$_{64}$+MHA$_8$)-LSTM$_{32}$ "0.7 D" | 94.18 | 0.1290 | 0.0106 | 99.31 | 0.0166 | 0.0013 |
| #12 | (BiGRU$_{64}$+MHA$_8$)-LSTM$_{32}$ "0.2 D" | 94.06 | 0.1221 | 0.0623 | 99.20 | 0.0211 | 0.0020 |

Abbreviations: Dropout (D)

For instance, L2D2 [18] achieves 98% accuracy on CICIoMT2024, but omits loss, FPR, and inference time. Similarly, BiGRU–LSTM in [22] reaches 98.32% accuracy on EdgeIIoT, with no F1-score, loss, or runtime reported. The DNN model from [39] records an even lower accuracy of 96.01%, while providing no supporting metrics. Likewise, the XGBoost model in [42] achieves only 95.01% on CICIoMT2024 and lacks any further evaluation criteria. The absence of FPR, inference time, and loss in all these studies limits their applicability in real-time, resource-constrained environments such as healthcare and industrial IoT, where both detection quality and operational efficiency are critical. On the other hand, three notable works—[16], [17], and [43]—conduct cross-domain testing across datasets representing different IoT contexts. However, their generalisation capability remains limited. For example, [43] achieves 99.88% on CICIoMT2024 but drops drastically to 33.30% on EdgeIIoT. Similarly, [16] and [17] report imbalanced precision-recall and omit runtime metrics, hindering deployment potential.

In contrast, the proposed BiGAT-ID model offers consistent and robust performance across domains, achieving up to 99.13% accuracy on CICIoMT2024 and 99.34% on EdgeIIoT, with exceptionally low FPRs (0.0013%), and inference times as low as 0.0002s and 0.0001s per instance, respectively. Notably, it is the only model to report a full set of evaluation metrics with stable performance across domains, highlighting its readiness for deployment in diverse and dynamic IoT environments.

To validate the generalizability of the proposed BiGAT-ID model, experiments were extended to the TON_IoT dataset, which comprises 10 classes of diverse IoT/IIoT attack types with varying traffic characteristics. Two pre-processing settings were applied: SMOTE with RoS and focal loss. As shown in Table 6, the SMOTE+RoS configuration achieved the best results on TON_IoT, with 98.67% accuracy, 99.36% precision, 98.94% recall, and 99.15% F1-score, while focal loss yielded 98.47% accuracy, 98.63% precision, 98.47% recall, and 98.51% F1-score. These results demonstrate the proposed BiGAT-ID model's robustness across datasets with varying characteristics and imbalance levels, showing superior performance on IIoT and IoMT datasets and maintaining competitive results on a general IoT dataset in terms of multiple DL performance metrics and inference time.

## 5. Conclusion

This paper introduced BiGAT-ID model, designed for intrusion detection across heterogeneous IoT environments. Validated on medical and industrial datasets, the model demonstrated consistent and high performance, aided by tailored balancing strategies to address class imbalance. Unlike previous approaches, BiGAT-ID provides a complete evaluation, including precision, latency, and FPR, confirming its readiness for real-time deployment in critical infrastructures. The study also emphasizes the importance of cross-domain generalization, an area where many existing models struggle. BiGAT-ID addresses this through architectural synergy between temporal modeling and attention mechanisms, resulting in stable outcomes across diverse IoT contexts. Given the limited availability of labeled IoT data, particularly in IoMT, future work may focus on leveraging transfer learning, enhancing interpretability, and optimizing the architecture for scalable real-world deployment. In particular, fine-tuning small language models could be explored to improve the efficiency of the proposed BiGAT-ID architecture across diverse IoT domains.

**Table 6**
Performance metrics of the proposed BiGAT-ID model in comparison to state-of-the-art methods for multiclass classification.

| Work | Model | Dataset | Acc. (%) | Loss | Pr (%) | Rc (%) | F1 (%) | FPR (%) | Inf time (sec/inst) |
|---|---|---|---|---|---|---|---|---|---|
| [18] | L2D2 (LSTM) | CICIoMT2024 | 98 | ✗ | 98 | 98 | 98 | ✗ | ✗ |
| [22] | BiGRU-LSTM | EdgeIIoT | 98.32 | ✗ | 98.78 | 97.22 | ✗ | ✗ | ✗ |
| [39] | DNN | EdgeIIoT | 96.01 | ✗ | ✗ | ✗ | ✗ | ✗ | ✗ |
| [42] | XGBoost | CICIoMT2024 | 95.01 | ✗ | ✗ | ✗ | ✗ | ✗ | ✗ |
| [16] | DNN | CICIoMT2024 | 99.56 | ✗ | 94.59 | 91.28 | 92.62 | ✗ | ✗ |
|  |  | CICIoT2023 | 99.09 | ✗ | 91.56 | 66.98 | 69.86 | ✗ | ✗ |
| [17] | Att-DNN | CICIoMT2024 | ✗ | ✗ | 84.43 | 98.73 | 91.02 | ✗ | ✗ |
|  |  | MQTT-IoT-IDS | ✗ | ✗ | 92.14 | 99.17 | 95.53 | ✗ | ✗ |
| [43] | CNN-LSTM-ResNet-SA | CICIoMT2024 | 99.88 | ✗ | 99.89 | 99.99 | 99.94 | ✗ | ✗ |
|  |  | EdgeIIoMT | 33.30 | ✗ | 33.31 | 100 | 49.97 | ✗ | ✗ |
| Our | BiGRU+MHA-LSTM (SMOTE+RoS) | CICIoMT2024 | 99.13 | 0.0257 | 99.13 | 99.13 | 99.13 | 0.0013 | 0.0002 |
|  |  | EdgeIIoT | 99.34 | 0.0158 | 99.34 | 99.34 | 99.34 | 0.0013 | 0.0001 |
|  |  | TON_IoT | 98.67 | 0.0476 | 98.68 | 98.67 | 98.67 | 0.0086 | 0.00016 |
|  | BiGRU+MHA-LSTM (Focal loss) | CICIoMT2024 | 94.98 | 0.0145 | 94.98 | 93.66 | 94.23 | 0.0293 | 0.00026 |
|  |  | EdgeIIoT | 99.13 | 0.0016 | 99.13 | 99.13 | 99.13 | 0.0013 | 0.00017 |
|  |  | TON_IoT | 98.47 | 0.0054 | 98.63 | 98.47 | 98.51 | 0.0068 | 0.00015 |

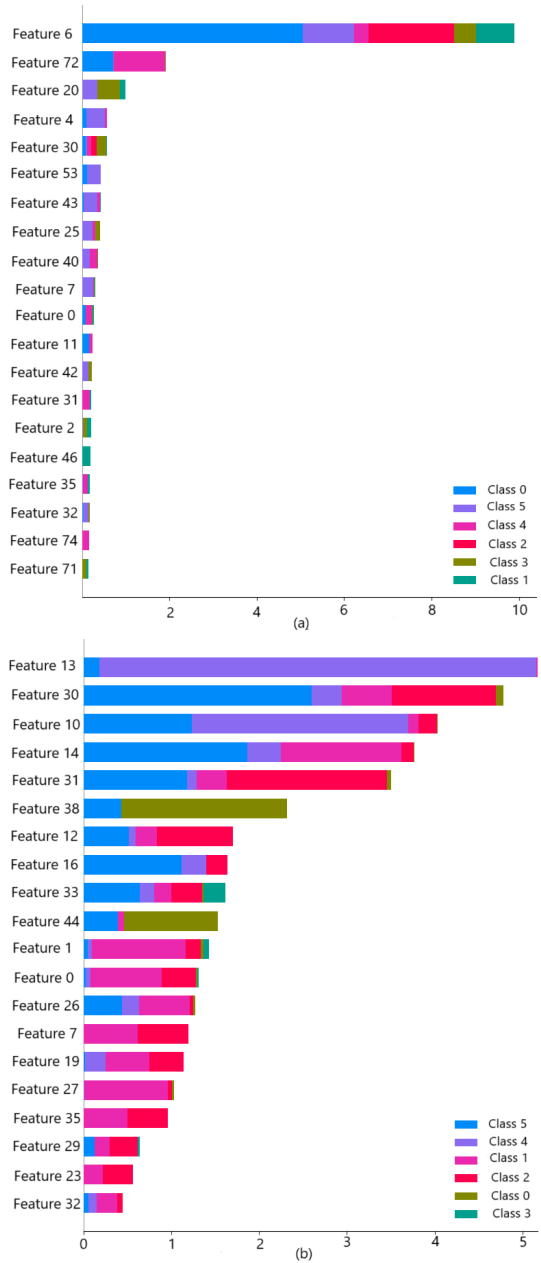

**Figure 5:** Mean SAHAP value average impact on model output magnitude. (a): For the CICIoMT2024 dataset; (b): For the EdgeIIoT dataset.

## Acknowledgment

The authors acknowledge that the Open Access APC was funded by the University of Liverpool (Grant No.).

## References

1. Ali Ghubaish, Tara Salman, Maede Zolanvari, Devrim Unal, Abdulla Al-Ali, and Raj Jain. Recent advances in the internet-of-medical-things (IoMT) systems security. *IEEE Internet of Things Journal*, 8(11): 8707–8718, 2020.
2. Hamza Kheddar, Yassine Himeur, and Ali Ismail Awad. Deep transfer learning for intrusion detection in industrial control networks: A comprehensive review. *Journal of Network and Computer Applications*, 220:103760, 2023.
3. Mohammed AM Farzaan, Mohamed Chahine Ghanem, Ayman El-Hajjar, and Deepthi N Ratnayake. Ai-powered system for an efficient and effective cyber incidents detection and response in cloud environments. *IEEE Transactions on Machine Learning in Communications and Networking*, 2025.
4. Fatma Yasmine Loumachi, Mohamed Chahine Ghanem, and Mohamed Amine Ferrag. Advancing cyber incident timeline analysis through retrieval-augmented generation and large language models. *Computers*, 14(2):67, 2025.
5. Hoe Tung Yew, Ming Fung Ng, Soh Zhi Ping, Seng Kheau Chung, Ali Chekima, and Jamal A Dargham. IoT based real-time remote patient monitoring system. In *2020 16th IEEE international colloquium on signal processing & its applications (CSPA)*, pages 176–179. IEEE, 2020.
6. Vangelis Malamas, Fotis Chantzis, Thomas K Dasaklis, George Stergiopoulos, Panayiotis Kotzanikolaou, and Christos Douligeris. Risk assessment methodologies for the internet of medical things: A survey and comparative appraisal. *IEEE Access*, 9:40049–40075, 2021.
7. Mohammad Alshinwan, Abdul Ghafoor Memon, Mohamed Chahine Ghanem, and Mohammed Almaayah. Unsupervised text feature selection approach based on improved prairie dog algorithm for the text clustering. *Jordanian Journal of Informatics and Computing*, 2025(1): 27–36, 2025.

8. Animesh Singh Basnet, Mohamed Chahine Ghanem, Dipo Dunsin, Hamza Kheddar, and Wiktor Sowinski-Mydlarz. Advanced persistent threats (apt) attribution using deep reinforcement learning. *Digital Threats: Research and Practice*.

9. Mikiyas Alemayehu, Mohamed Chahine Ghanem, Karim Ouazzane, Hamza Kheddar, and Marcio J Lacerda. A systematic analysis on the use of ai techniques in industrial IoT ddos attacks detection, mitigation and prevention. 2025.

10. Mohamed Ghanem, Fadi Dawoud, Habiba Gamal, Eslam Soliman, Tamer El-Batt, and Hossam Sharara. Flobc: A decentralized blockchain-based federated learning framework. In *2022 Fourth International Conference on Blockchain Computing and Applications (BCCA)*, pages 85–92. IEEE, 2022.

11. Mohamed Ghanem, Abdelaaziz Mouloudi, and Mohammed Mourchid. Towards a scientific research based on semantic web. *Procedia Computer Science*, 73:328–335, 2015.

12. Hamza Kheddar, Diana W Dawoud, Ali Ismail Awad, Yassine Himeur, and Muhammad Khurram Khan. Reinforcement-learning-based intrusion detection in communication networks: A review. *IEEE Communications Surveys & Tutorials*, 2024.

13. Mohamed Chahine Ghanem and Deepthi N Ratnayake. Enhancing wpa2-psk four-way handshaking after re-authentication to deal with de-authentication followed by brute-force attack a novel re-authentication protocol. In *2016 International Conference On Cyber Situational Awareness, Data Analytics And Assessment (CyberSA)*, pages 1–7. IEEE, 2016.

14. Hamza Kheddar. Transformers and large language models for efficient intrusion detection systems: A comprehensive survey. *Information Fusion*, page 103347, 2025.

15. Alireza Mohammadi, Hosna Ghahramani, Seyyed Amir Asghari, and Mehdi Aminian. Securing healthcare with deep learning: A CNN-based model for medical IoT threat detection. In *2024 19th Iranian Conference on Intelligent Systems (ICIS)*, pages 168–173. IEEE, 2024.

16. Duc Vu-Minh, My Duong-Tran-Tra, Luan Van-Thien, Anh Pham-Nguyen-Hai, Thuat Nguyen-Khanh, and Quan Le-Trung. Performance evaluation of decentralized machine learning based network-based intrusion detection system for internet of things. In *2024 18th International Conference on Advanced Computing and Analytics (ACOMPA)*, pages 78–85. IEEE, 2024.

17. Mireya Lucia Hernandez-Jaimes, Alfonso Martinez-Cruz, Kelsey Alejandra Ramírez-Gutiérrez, and Alicia Morales-Reyes. Network traffic inspection to enhance anomaly detection in the internet of things using attention-driven deep learning. *Integration*, page 102398, 2025.

18. Gökhan Akar, Shaaban Sahmoud, Mustafa Onat, Ünal Cavusoglu, and Emmanuel Malondo. L2D2: a novel LSTM model for multi-class intrusion detection systems in the era of IoMT. *IEEE Access*, 2025.

19. Jiajun Zhou, Wentao Fu, Hao Song, Shanqing Yu, Qi Xuan, and Xiaoniu Yang. Multi-view correlation-aware network traffic detection on flow hypergraph. *arXiv preprint arXiv:2501.08610*, 2025.

20. Amina Khacha, Rafika Saadouni, Yasmine Harbi, and Zibouda Aliouat. Hybrid deep learning-based intrusion detection system for industrial internet of things. In *2022 5th International Symposium on Informatics and its Applications (ISIA)*, pages 1–6. IEEE, 2022.

21. Afrah Gueriani, Hamza Kheddar, and Ahmed Cherif Mazari. Adaptive cyber-attack detection in iiot using attention-based LSTM-CNN models. In *2024 International Conference on Telecommunications and Intelligent Systems (ICTIS)*, pages 1–6. IEEE, 2024.

22. Danish Javeed, Tianhan Gao, Muhammad Shahid Saeed, and Prabhat Kumar. An intrusion detection system for edge-envisioned smart agriculture in extreme environment. *IEEE Internet of Things Journal*, 2023.

23. Rafika Saadouni, Amina Khacha, Yasmine Harbi, Chirihane Gherbi, Saad Harous, and Zibouda Aliouat. Secure iiot networks with hybrid cnn-gru model using edge-iiotset. In *2023 15th International Conference on Innovations in Information Technology (IIT)*, pages 150–155. IEEE, 2023.

24. Dusmurod Kilichev, Dilmurod Turimov, and Wooseong Kim. Next–generation intrusion detection for IoT evcs: Integrating cnn, lstm, and gru models. *Mathematics*, 12(4):571, 2024.

25. Kai Yang, JiaMing Wang, and MinJing Li. An improved intrusion detection method for iiot using attention mechanisms, bigru, and inception-cnn. *Scientific Reports*, 14(1):19339, 2024.

26. *Towards an efficient automation of network penetration testing using model-based reinforcement learning*. PhD thesis, City, University of London, 2022.

27. Mohamed C Ghanem and Said Salloum. Integrating ai-driven deep learning for energy-efficient smart buildings in internet of thing-based industry 4.0. *Babylonian Journal of Internet of Things*, 2025:121–130, 2025.

28. Manish Kumar, Changjong Kim, Yongseok Son, Sushil Kumar Singh, and Sunggon Kim. Empowering cyberattack identification in ioht networks with neighborhood component-based improvised long short-term memory. *IEEE Internet of Things Journal*, 2024.

29. Manish Kumar and Sunggon Kim. Securing the internet of health things: Embedded federated learning-driven long short-term memory for cyberattack detection. *Electronics*, 13(17):3461, 2024.

30. Easa Alalwany, Bader Alsharif, Yazeed Alotaibi, Abdullah Alfahaid, Imad Mahgoub, and Mohammad Ilyas. Stacking ensemble deep learning for real-time intrusion detection in IoMT environments. *Sensors*, 25(3):624, 2025.

31. Zhen Wang, Anazida Zainal, Maheyzah Md Siraj, Fuad A Ghaleb, Xue Hao, and Shaoyong Han. An intrusion detection model based on convolutional kolmogorov-arnold networks. *Scientific Reports*, 15(1): 1917, 2025.

32. Mohammed Mudassir, Devrim Unal, Mohammad Hammoudeh, and Farag Azzedin. Detection of botnet attacks against industrial IoT systems by multilayer deep learning approaches. *Wireless Communications and Mobile Computing*, 2022(1):2845446, 2022.

33. Safi Ullah, Wadii Boulila, Anis Koubaa, and Jawad Ahmad. MAGRU-IDS: a multi-head attention-based gated recurrent unit for intrusion detection in iiot networks. *IEEE Access*, 2023.

34. Mohammed S Alshehri, Oumaima Saidani, Fatma S Alrayes, Saadullah Farooq Abbasi, and Jawad Ahmad. A self-attention-based deep convolutional neural networks for iiot networks intrusion detection. *IEEE Access*, 2024.

35. Alberto Fernández, Salvador Garcia, Francisco Herrera, and Nitesh V Chawla. Smote for learning from imbalanced data: progress and challenges, marking the 15-year anniversary. *Journal of artificial intelligence research*, 61:863–905, 2018.

36. Yesi Novaria Kunang, Siti Nurmaini, Deris Stiawan, and Bhakti Yudho Suprapto. Deep learning with focal loss approach for attacks classification. *TELKOMNIKA (Telecommunication Computing Electronics and Control)*, 19(4):1407–1418, 2021.

37. Afrah Gueriani, Hamza Kheddar, and Ahmed Cherif Mazari. Enhancing IoT security with cnn and lstm-based intrusion detection systems. In *2024 6th International Conference on Pattern Analysis and Intelligent Systems (PAIS)*, pages 1–7. IEEE, 2024.

38. Sajjad Dadkhah, Euclides Carlos Pinto Neto, Raphael Ferreira, Reginald Chukwuka Molokwu, Somayeh Sadeghi, and Ali Ghorbani. Ciciomt2024: Attack vectors in healthcare devices-a multi-protocol dataset for assessing iomt device security, 2024.

39. Mohamed Amine Ferrag, Othmane Friha, Djallel Hamouda, Leandros Maglaras, and Helge Janicke. Edge-iiotset: A new comprehensive realistic cyber security dataset of IoT and IIoT applications for centralized and federated learning. *IEEE Access*, 10:40281–40306, 2022.

40. Abdullah Alsaedi, Nour Moustafa, Zahir Tari, Abdun Mahmood, and Adnan Anwar. Ton_iot telemetry dataset: A new generation dataset of iot and iiot for data-driven intrusion detection systems. *Ieee Access*, 8: 165130–165150, 2020.

41. Scott M Lundberg and Su-In Lee. A unified approach to interpreting model predictions. *Advances in neural information processing systems*, 30, 2017.

42. Fatima Sohail, Muhammad Asim Mukhtar Bhatti, Muhammad Awais, and Aamna Iqtidar. Explainable boosting ensemble methods for intrusion detection in internet of medical things (IoMT) applications. In *2024 4th International Conference on Digital Futures and Transformative Technologies (ICoDT2)*, pages 1–8. IEEE, 2024.

43. Tinshu Sasi, Arash Habibi Lashkari, Rongxing Lu, Pulei Xiong, and Shahrear Iqbal. An efficient self attention-based 1D-CNN-LSTM network for IoT attack detection and identification using network traffic. *Journal of Information and Intelligence*, 2024.